\documentclass[prl,showpacs, twocolumn,groupedaddress]{revtex4}

\usepackage{graphicx}

\begin{document}

\title
{Spin and temperature dependent study of
 exchange and correlation in thick two-dimensional
 electron layers. 
 }

\author
{ M.W.C. Dharma-wardana
}
\email[Email address:\ ]{chandre.dharma-wardana@nrc.ca}
\affiliation{
Institute of Microstructural Sciences,
National Research Council of Canada, Ottawa, Canada. K1A 0R6\\
}
\date{\today}
\begin{abstract}
The exchange and correlation $E_{xc}$ of strongly correlated
electrons in 2D layers of finite width are studied as a function
of the density parameter $r_s$, spin-polarization $\zeta$
 and the temperature $T$.
We explicitly  treat strong-correlation effects via
pair-distribution functions, and introduce an equivalent
constant-density approximation (CDA) applicable to all the inhomogeneous
densities encountered here.
The width $w$ defined via the CDA provides a length scale defining the
$z$-extension of the quasi-2D layer resident in the $x$-$y$ plane.
The correlation energy $E_c$ of the quasi-2D system is presented 
as an interpolation between a 1D gas of electron-rods (for
$w/r_s>1$) coupled via a log(r) interaction, and a 3D Coulomb fluid 
closely approximated from the known {\it three-dimensional}
 correlation energy when $w/r_s$ is small. 
Results for the $E_{xc}(r_s,\zeta,T)$, the 
transition to a spin-polarized phase,
the effective mass $m^*$, the Land\'e $g$-factor etc.,
 are reported here.
\end{abstract}
\pacs{PACS Numbers: 05.30.Fk, 71.10.+a, 71.45.Gm}
%
\maketitle
\section{Introduction.}
The
 2D electron systems (2DES) present in GaAs or Si/SiO$_2$ structures
 access a
wide range of electron densities,
providing a wealth of experimental observations\cite{krav}.
The 2D electrons reside in the x-y plane and also
have a transverse ($z$-dependent)
density $n(z)$ which is
 confined to the lowest subband of the
hetero-structure\cite{afs}. The 2D character arises 
since the higher subbands
 are 
sufficiently  above the Fermi energy $E_F$.
Then the physics depends only on the
  ``coupling parameter''
$\Gamma$ = (potential energy)/(kinetic energy), the
$z$-distribution $n(z)$, the spin-polarization $\zeta$,
and the temperature $T$ which has to be
 significantly smaller than the Fermi energy
$E_F$ to preserve the 2-D character. 
The $\Gamma$ for the 2DES at the density $n$ is
equal to the mean-disk radius $r_s=(\pi n)^{-1/2}$ per electron,
expressed in effective atomic units which
depend on the bandstructure mass $m_b$ and the ``background''
dielectric constant $\epsilon_b$.
The $z$-motion in the lowest subband
 may have widths of  $\sim 600$ \AA, in GaAs when $r_s$ is  $\sim 6$, in
 heterojunction insulated-gate field-effect transistors (HIGFET)
which have been an object of recent studies\cite{zhu}. 
Similarly, other nanostructures (e.g., 
quantum dots) contain electrons confined to a micro-region in
the $x$-$y$ plane, and have a sizable  $z$ extension\cite{williams}. 
 Hence layer-thickness effects are important 
 in many areas of nanostructure physics.
Appropriate correlation
functionals\cite{martin} for such systems are still unavailable, even though
 the exchange functional for Fang-Howard distributions is known
\cite{sternjjap}.

Layer thickness effects are a long standing
probe of exchange and
correlation theories in 3D electron slabs\cite{perdew}.
The relevance of the finite size of the 2D layers had
also been considered within the
quantum Hall effect\cite{ahm, dassarmaqh}, and more recently 
in the context of the $g$-factor and the effective mass $m^*$ of
2-D layer\cite{tutuc,tan,morsen,asgari,zhang,cdw}.
 In the early days of the application of 
diagrammatic perturbation theory (PT) to condensed-matter problems,
it was normal to attempt to calculate various many-body
properties like the effective mass $m^*$, the effective $g$-factor
$g^*$, and corrections to the total energy using perturbation
methods. The need to go beyond the random-phase approximation (RPA)
was rapidly appreciated and lead to the work of Hubbard, Rice,
Vosko, Geldart and others\cite{geldart}.
The common experience with the generalized RPA method, when
 applied to 3D  electrons and  to ideally thin
2D layer is that it predicts a spin-polarized phase at unrealistic
high densities (low $r_s$), while Quantum Monte Carlo
(QMC) simulations suggest a density near $r_s\sim$ 25-26 in 2D.
RPA methods lead to negative pair-distribution functions (PDFs),
incorrect local-field corrections in the response functions,
and disagreement with the compressibility sum rule
and other formal conditions. Recent attempts to calculate the effective
mass $m^*$ for ideally thin layers\cite{asgari,zhang} show strong
disagreement with the $m^*$ obtained from
 QMC simulations\cite{kwon}.
 In fact, the main thrust of
the programs of 
Singwi,
Tosi et al. (STLS)\cite{stls}, and Ichimaru et al.\cite{ichimaru}
was to overcome 
the shortcomings of the RPA-like approach via
non-perturbative methods. 
Many-body calculations where different parts of the 
calculation (e.g, vertex corrections, local-field corrections,
etc.) are obtained from different sources, (e.g, 
vertex corrections from a model, local-field corrections
or correlation energies from QMC,
and  some other quantities  fitted to sum rules etc.)
and combined together, have also appeared.
 Unless the same ``mixture'' is
used to calculate a multitude of properties and shown to lead to
consistent results, such theories have to be treated with caution.

\begin{figure}
\includegraphics*[width=9.0cm, height=10.0cm]{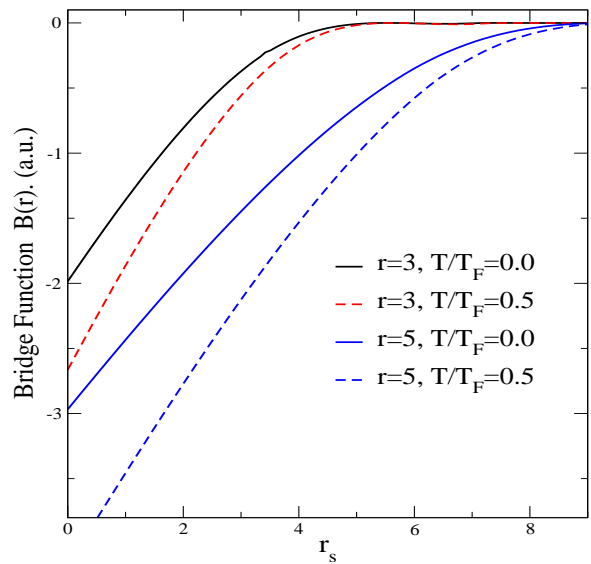}
\caption
{The bridge function $B(r)$ for $r_s=3$ and $r_s=5$. The effect of
changing the temperature is also shown. 
}
\label{bridge}
\end{figure}

 We have
 introduced an approximate
  method for strongly correlated quantum systems
where the objective is to work with the PDF of the quantum
fluid, generated from an equivalent
 classical Coulomb fluid whose temperature $T_q$ is chosen to
reproduce the correlation energy of the original quantum fluid at $T=0$.
The classical PDFs are calculated via the hyper-netted-chain (HNC)
equation, and the method is called the CHNC.
As the method has been described
 in previous publications\cite{prl1,prb,prl2,prl3},
 and successfully applied to a variety 
 of problems\cite{hyd,2valley,locf,eplmass}, 
only a brief account is given here, mainly to 
help the reader.
The PDFs are obtained from an integral equation
which can be recast into a classical Kohn-Sham
form where the correlation effects are captured as a sum of
HNC diagrams and bridge diagrams. 
\begin{equation}
\label{chnceq}
g_{ij}(r)=\exp^{-\beta\left(P_{ij}(r)+V_{cou}(r)+ N_{ij}(r)+B_{ij}(r)\right)}
\end{equation}
The temperature of the classical fluid $T_q=1/\beta$ is chosen such that
at $T=0$ the calculated classical $g(r)$
 recovers the known correlation energy
of the fully spin-polarized 2D electron fluid at the given density.
This fitting has been done in ref.~\cite{prl2}, and $T_q$ is known
as a function of $r_s$. At finite-$T$, the classical fluid temperature
$T_{cf}$ is taken to be 
\begin{equation}
T_{cf}=(T^2_q+T^2)^{1/2}
\end{equation}
This $T_{cf}$ is used for all spin polarizations.
In Eq.~\ref{chnceq} the indices
 $i,j$ label the spin states. The $P_{ij}(r)$ is chosen
to ensure that $g_{ij}(r)$ reduces to the
explicitly known non-interacting PDF, $g_{ij}^0(r)$,
when the Coulomb interaction $V_{cou}(r)$ is switched off. Thus
$P_{ij}(r)$ takes care of Pauli-exclusion effects and ensures that the
``Fermi-hole'' is exactly recovered\cite{lado}. Also, $ N_{ij}(r)$ is a sum of
terms that appear in the hyper-netted-chain 
equation, while $B_{ij}(r)$ contains 3-body and higher-order diagrams
not contained in the nodal term  $N_{ij}(r)$. The latter depends
implicitly on $g_{ij}(r)$, and is evaluated via the Ornstein-Zernike
equation.  The bridge term is very difficult to evaluate, but 
the hard-sphere fluid 
provides a good approximation. That is, in 2D, we specify
$B_{ij}(r)$ by specifying the  hard-disk radius $r_{H}$, or equivalently
 the packing fraction $\eta$, and use the Percus-Yevik
approach\cite{harddisk}. As the
parallel-spin three-body clusters are
suppressed by the Pauli exclusion,
 we use only a single 
bridge function, viz., $B_{12}(r)$. This makes the bridge interaction
effectively independent of $\zeta$. Bulutay and Tanatar, and
also Khanh and Totsuji\cite{buluty}, have examined variants 
of CHNC without a bridge function (this is some what equivalent
to neglecting back-flow terms in QMC simulations of 2D systems).
 The hard-sphere
radius $r_H$, and the packing fraction $\eta$
 are given by\cite{prl2}:
\begin{eqnarray}
r_H&=&2r_s\eta^{1/2}\\
\eta&=&0.1175r_s^{1/3}(t_{cf}+t^2/2)\\
t_{cf}&=&T_{cf}/E_F,\;\;\;t=T/E_F
\end{eqnarray}
A plot of the bridge function  for a few typical cases is given
in Fig.~\ref{bridge}. 
%

The
 CHNC method was applied to the 3D and
2D electron fluids\cite{prl1,prb,prl2,prl3},
 to dense hydrogen fluid\cite{hyd}, 
and also to the two-valley system in Si/SiO$_2$
2DES\cite{2valley, eplmass}. 
In each case we showed that the 
PDFs, energies
 and other properties obtained from CHNC were in excellent
agreement with comparable QMC results.
The advantage of the CHNC method is that it affords a simple, semi-analytic
theory for strongly correlated systems where QMC becomes prohibitive
or technically impossible to carry out. The classical-fluid model
allows for physically motivated treatments of complex issues
like three-body clustering etc., which are
difficult within quantum methods.
The disadvantage, 
typical of such many-body approaches, is that at present it
remains an ``extrapolation'' taking off from the results of  
a model fluid. The fully spin-polarized
infinitely-thin uniform 2DES is the model fluid\cite{prl2}, as the
$E_c$  of this one-component system is  accurately known.

In this study  we  use  a method of replacing the inhomogeneous electron
distribution $n(\vec{r},z)=n(\vec{r})n(z)$ by a homogeneous density
 system\cite{ggsavin,cdw}
i.e., a constant-density approximation (CDA)
 where the transverse density $n(z)$ is a
constant within a slab of width $w$, and zero outside.
 The CDA avoids difficulties associated with
gradient expansions noted in ref.\cite{martin}. Also, the CDA presents a
unified approach to quasi-2D distributions like the
Fang-Howard model\cite{afs}, the quantum-well model etc.
$E_{xc}$ for such distributions
is calculated as a function of the spin polarization $\zeta$, 
2-D density parameter $r_s$, the CDA width $w$
 and the temperature $T$.
This enables us to determine physical quantities
related to the Landau Fermi liquid parameters.
Thus the spin susceptibility enhancement,
 the effective mass $m^*$, and
 the Land\'{e} $g$ factor for the quasi 2D electrons
are presented.
\section{The quasi-2D interaction}
The transverse distribution $n(z)$  of the 2D electrons is given by the 
square of the lowest subband wavefunction $\phi(z)$ of the heterostructure,
calculated within the envelope approximation. The nature of the materials used
(e.g, Si/SiO$_2$ or GaAs) and
the doping profiles determine the confining potential
and the electron density $n(z)$ in the z-direction.
Typically, $n(z)$  may be modeled by one of the
 following forms.
\begin{eqnarray}
\label{distributions}
n(z)&=&\delta(z), \;\;\; \mbox{ideal thin layer}  \\
       &=&(2/w)\sin^2(\pi z/w), \;\;\; \mbox{infinite well}\\
       &=&(b^3/2)z^2\exp(-bz),\; \;\; \mbox{Fang-Howard} \\
       &=&1/w,\;\; |z|\le w    \;\;\; \mbox{constant-density model}    
\end{eqnarray}
The second  and third 
are frequently used  approximate (but adequate)  models, while
we present the fourth model, the CDM. This is an excellent
constant-density  approximation
(to be called the constant-density approximation, CDA)
to generate the effective 2-D potential
 arising from
most $n(z)$ distributions, on suitably choosing $w$.
The Fang-Howard (FH) form\cite{afs,fangh},
contains the parameter $b$, and is
normalized in the range $0$ to $\infty$.
The FH parameter $b$ is such that
 $b^3=(48*\pi/a_0)(n_d+11n_s/32)$,
 where $n_s$ is the 2D electron density $n$,
 and $n_d$ is the
depletion charge density\cite{afs}. The material parameters are
contained in the effective Bohr radius
 $a_0=\epsilon_b\hbar^2/m_b e^2$ defined in terms of 
the usual constants, $\epsilon_b$ and $m_b$ being the
 dielectric constant and the
band mass.
For the devices used in ref.~\cite{tan,zhu},
the depletion density has been reported to be negligible\cite{morsen}.
Then $b^3=33/(2r_s^2)$, in atomic units.
\subsection{The constant-density model.}
\begin{figure}
\includegraphics*[width=9.0cm, height=10.0cm]{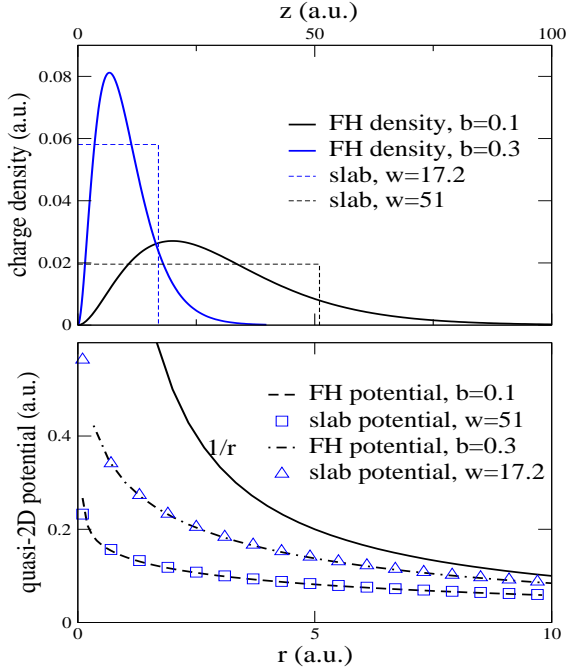}
\caption
{(a)The Fang-Howard density and the equivalent constant-density slab.
(b)Comparison of the $1/r$ form with the quasi-2D potential from
 the Fang-Howard distribution
 for the FH parameter $b$ = 
0.1, and 0.3 and the equivalent slabs at $w$ =  53.3 and 17.8 a.u. 
respectively.}
\label{pots}
\end{figure}

We denote the Coulomb potential in an infinitely thin layer by
$V(r)=1/r$, while $W(r)$ is used for the effective 2-D
potential of a thick layer.
The effective 2D-Coulomb potential  $W(r)$
 between two electrons having coordinates ($\vec{r_1}, z_1$)
and ($\vec{r_2}, z_2$), with $\vec{r}=\vec{r_1}-\vec{r_2}$ is given by,
\begin{equation}
W(r)=\int_0^{z_{m}}\int_0^{z_{m}}\frac{dz_1dz_2 n(z_1)n(z_2)}
{[r^2+(z_1-z_2)^2]^{1/2}}\\
\end{equation}
Here $z_{m}$ is $\infty$ for FH, while $z_{m}=w$ for the others. 
The potential $W(r)=(1/r)F(r)$ and the form factor $F(r)$
reflects the effect of the $z$-extension of the density.
There is no analytic form for $F(r)$ in the Fang-Howard
case, while the $q$-space form, $F(q)$ is known\cite{afs}.
If the dielectric constants of the barrier and well material
were assumed equal, then the Fang-Howard form factor is:
\begin{equation}
F(q)=[1+\frac{9q}{8b}+\frac{3q^2}{8b^2}][1+\frac{q}{b}]^{-2}
\end{equation}
Here we derive a potential $W(r,w)$ for the 
constant-density model(CDM) which is, to an excellent approximation
 {\it electrostatically equivalent} to the
the 2D potential  for any reasonable $n(z)$.
 These FH-type distributions are themselves
convenient fits to the self-consistent Schrodinger solutions and
have uncertainties of a few percent. The
CDA holds well within such limits. The potentials are 
explicitly shown in Fig.~\ref{pots} for the FH form
where we have taken an extreme example with $b=0.1$.
The  method of replacing an inhomogeneous distribution by a
uniform distribution is suggested by the observation that the 
non-interacting total pair-correlation function $h^0(r)$ has
 the form $\sim n(r)^2$,
where $n(r)$ is the density
-profile around the Fermi hole.
In our case we wish to replace the inhomogeneous $n(z)$ by a
constant-density distribution $n_{cd} $ which has 
 the same  electrostatic potential
in the 2-D plane as $n(z)$. 
\begin{equation}
n_{cd}=1/w=\int n(z)^2dz
\end{equation}
Since the subband distribution is normalized to unity, the width
$w$ of the constant-density slab is simply $1/n_{cd}$.
Starting from different objectives, 
Gori-Giorgi et al.\cite{ggsavin}
have already proposed this formula for determining
an average density for an inhomogeneous density, in the
context of pair-distributions in 2-electron atoms.
We have also shown the utility of the CDA in estimating
the correlation energy of the 2DES in the Wigner-crystal
phase\cite{jost}.
The $w$ of the CDA is somewhat different from the
``thickness'' $3/b$ often assigned to the FH distribution.
In fact, the constant-density slab width $w$
for the Fang-Howard $b$ is given by $w=16/(3b)$. 
The quasi-2D potential for a CDM
 of width $w$ is given by
\begin{eqnarray}
\label{cdmeqn}
W(r)&=&V(r)F(s),\;\; s=r/w, \;\; t=\surd{(1+s^2)} \\
F(s)&=&2s\left[\log\frac{1-t}{s}
\,\,+1-t\right] 
\end{eqnarray}
This potential tends to $1/r$ for large $r$, and behaves as
$$\frac{2}{w}\left(\ln\frac{2w}{r}+\frac{r}{w}-1\right)$$
for $r < w$. Thus the short-range behaviour is logarithmic
and  weaker than the
 Coulomb potential.
The
$k$-space form of the CDM potential is:
\begin{eqnarray}
V_{usm}(k,w)&=&V(k)F(p),\,\,\, p=kw\\ 
F(p)&=&(2/p)\{(e^{-p}-1)/p+1\}
\end{eqnarray}
The form factors $F(s)$ and $F(p)$ tend to unity as $w\to 0$.
These $r$-space and $k$-space analytic forms of the CDM  
lead to analytic formulae for the FH form.
In our work we assume that a given distribution has been
replaced by an equivalent uniform-slab distribution, and only the
final $W(r)$ potential enters into the exchange-correlation
calculations (the numerical work has been checked
via direct calculations as well).
 In the case of GaAs-HIGFETS, if $n_d$ could be
neglected, the $r_s$ parameter
specifies the $b$ parameter and hence the width $w$ of the CDM. 
Then $b\sim r_s^{-2/3}$ and $w=2.09494r_s^{2/3}$.
\section{Exchange Free Energy for quasi-2D layers.}
\label{expara}
The exact exchange free energy $F_x$ for 2D layers of finite thickness
 can be readily 
evaluated using the quasi-2D potential $W(r)$ and the noninteracting
pair-distribution functions $g^0_{ij}(r)$ of the 2-D fluid.
 Only the parallel-spin
case $i=j$ is relevant. Also, $g^0_{ij}(r)$ for a slab of finite
 thickness is
 identical
to that for an ideally thin 2D layer, both at $T=0$ and
 at finite-$T$. In fact,
we find that the $T$ dependence of the
$F_x$ of layers of
finite thickness is very close to that of the ideally thin case.
\subsection{Ideally-thin layer}
The first-order unscreened exchange free energy $F_x$ consists
of  $F_x^i$, where $i$ denotes the
two spin species. At $T=0$ these reduce to the exchange
energies:
\begin{equation}
E_i^x/n=\frac{8}{3\surd{\pi}}
n_i^{1/2}
\end{equation}
Here $n_1=n(1+\zeta)/2$, and $n_2=n(1-\zeta)/2$.
Then the exchange energy per particle at $T=0$, i.e., the total $E_x/n$
becomes
\begin{equation}
\label{ex_t=0}
E_x/n=(E_1^x+E_2^x)/n=-\frac{8}{3\pi r_s}[c_1^{3/2}+c_2^{3/2}]
\end{equation}
where $c_1$ and $c_2$ are the fractional compositions
$(1\pm\zeta)/2$ of the two spin species.

We define a reduced temperature $t=T/E_F$,
$E_F=\pi n$, and the species-dependent
reduced chemical potentials  $\mu^0_i/T$ by $\eta_i$,
 reduced
temperatures $t_1=t/(1+\zeta)$ and $t_2=t/(1-\zeta)$, based on the
two Fermi energies $E_{F1}$ and $E_{F2}$ which are $E_F(1\pm\zeta)$.
Then we have:
\begin{equation}
\label{exF}
F_i^x/E_i^x=\frac{3}{16}t_i^{3/2}\int_{-\infty}^{\eta_i}
\frac{I^2_{-1/2}(u)du}{(\eta_i-u)^{1/2}}
\end{equation}
The $I_{-1/2}$ is the Fermi integral defined as usual:
\begin{equation}
I_\nu(z)=\int_0^\infty\frac{dx x^\nu}{1+e^{x-z}}
\end{equation}
The $\eta_i$ are given by
\begin{equation}
\eta_i =\log(e^{1/t_i}-1)
\end{equation}
In the paramagnetic case
 Eq.~\ref{exF} reduces to the result given by
Isihara et al. (see their Eq.3.4; they use
a different definition of the Fermi
integral). For small values of $t$, the exchange energy is
of the form,
\begin{equation}
E_x(r_s,t)=E_x(r_s,0)[1+(\pi^2/16)t^2\log(t)-0.56736t^2+...]
\end{equation}
The total exchange free energy is $F_x=\Sigma F_i^x$.
The accurate numerical evaluation of Eq.~\ref{exF} requires the removal of
the square-root singularity by adding and subtracting, e.g,
$I^2(-|\eta|)/(v-|\eta|)^{1/2}$ for the case where $\eta$ is negative, and
$v=u$, and so on. 

 A real-space formulation of $F_x$ = $F_1^x+F_2^x$
using the zeroth-order PDFs fits  naturally with
the approach of our study. Thus
\begin{equation}
F_x/n=n\int \frac{2\pi r dr}{r}\sum_{i<j}h^0_{ij}(r)
\end{equation}
Here  $h^0_{ij}(r)=g^0_{ij}(r)-1$.
In the non-interacting system at temperature $T$,
 the antiparallel $h^0_{12}$, viz.,
 $g_{12}^0(r,T)-1$,
 is zero while
$$h_{11}^0({\bf r})
=-\frac{1}{n_i^2}\Sigma_{{\bf k}_1,{\bf k}_2}n(k_1)n(k_2)
e^{i({\bf k}_1-{\bf k}_2){\bf \cdot}{\bf r}} \,\,
=\, -[f(r)]^2$$
Here {\bf k}, {\bf r} are 2-D vectors and $n(k)$ is
 the Fermi occupation number at
the temperature $T$. At $T=0$ $f(r)=2J_1(k_ir)/k_r$ where
$J_1(x)$ is a Bessel function. As a numerical check, we have evaluated the
exchange free energy by {\it both} methods, i.e., via k-space and r-space
formulations.

We present a  convenient analytic fit to the exchange free energy which
is a universal function
$F_x(t)/E_x$, for arbitrary $\zeta$. That is, the  same function applies to
 any component, on using the
reduced Fermi temperature of the spin species.
The total $F_x$ is obtained by adding  both spin contributions.
The analytic fit is:
\begin{eqnarray}
\lefteqn{F_i^x(t,\zeta)/E_i^x(\zeta)=}\\
 &  &\frac{1+C_1t_i+C_2t_i^2}
{1+C_3t_i+C_4t_i^2}\,\tanh(1/\surd{t_i})\nonumber
\end{eqnarray}
The fit coefficients $C_i$ are
3.27603, 4.81484, 3.33100, 6.51436.
The temperature $t_i$ is $t/(1\pm\zeta)$, appropriate to the
spin polarization.
The  exchange effects in the 2DES decay
more slowly with temperature than in the 3D case where
 a  tanh$(1/t)$ factor
appears in Eq.~3.2 of ref.~\onlinecite{pdw84}. The above form does
not explicitly contain the low-temperature logarithmic term, but it
reproduces the value of 0.99382 at $t=0.05$, while the numerical
integration gives 0.9939497. Similarly, at $t=1$, 10 and 30
the fit (integral) returns 0.63839 (0.63839), 0.22999 (0.22990),
and 0.13421 (0.13410) respectively. 
\subsection{Thick 2-D layers.}
The $T=0$ exchange energy is modified by the layer thickness $w$.
The expression for $g^0_{ij}(r)$ depends only on
$x=r/r_s$. Similarly, the quasi-potential $W(r)$ depends on the 
reduced variable
$s=r/w$. Hence the exchange energy of the quasi-2D layer depends only on the
ratio $w_s=w/r_s$.
The exchange energy per electron at a density $n$,
 given by $r_s$, polarization
$\zeta$, in a layer of width $w$ is given by:
\begin{equation}
\label{exen}
Ei_x(r_s,\zeta,w)=\frac{1}{2}n r_s\int_0^\infty 2\pi dxh^0(x,\zeta)F(x/w_s)
\end{equation}
  Even though we have analytic forms for
$W(r)$ and $h^0(r)$ as well as their Fourier transforms, we have not found
a convenient analytic result for the exchange energy at $T=0$.
 However, the results can be
 parametrized by  simple analytic-fit formulae:
\begin{eqnarray*}
\label{exparam}
E_x(r_s,\zeta, w)&=&E_x(r_s,\zeta, 0)Q(w_s,\zeta),\,\, w_s=w/r_s\\
Q(w_s,\zeta)&=&\frac{1+A(\zeta)\surd{w_s}}{1+B(\zeta)\surd{w_s}+C(\zeta)w_s}\\
\end{eqnarray*}
Here $E_x(r_s,\zeta, 0)$ is the exchange energy of the ideally thin
system given by eq.~\ref{ex_t=0}. 
The ratio $Q(w_s,\zeta)$ is a measure of the reduction in
the exchange energy due to the thickness effect. Since the effect
depends on $w_s=w/r_s$, for a given thickness, 
the effect is greater for {\it high
density} samples. If the depletion density $n_d$ in HIGFETS
 could be neglected,
and if the exchange-correlation energy $E_{xc}$ is
 {\it not} included in the
energy minimization which determines the
 Fang-Howard parameter $b$, then
$w_s\sim 2.09r_s^{-1/3}$.
 The inclusion of $E_{xc}$ in self-consistently
 determining $b$ changes
$b$ by $\sim 2\%$ for low $r_s$, but the effect
 becomes less important at
higher $r_s$.
At $r_s$=1, and 10  for $\zeta=0$, the ratio  $Q$ is 0.652 and  0.794
 respectively.
The reduction from the ideally-thin 2D form is
clearly  substantial.
The exchange free energy $F_x(r_s,\zeta,T,w)$
 at finite-$T$, for layers with thickness $w$ is
found to be adequately approximated by the temperature factor of the
ideally-thin system. However, in calculating the effective mass
from the finite-$T$ energies, we make independent
 calculations near
$T=0$ and {\t do not} use the fit given here.

\begin{table}
\caption{
Parameters fitting the exchange energy ratio as a function of the layer width
ratio $w_s=w/r_s$.
}
\begin{ruledtabular}
\begin{tabular}{cccccc}
$\zeta$ & $A$ & $B$ & $C$ \\
\hline
0.0   &0.155294 &  0.142486  &  0.320735   \\
0.5   &0.184536 & 0.169822   &  0.370001 \\
1.0   &0.19838  & 0.179455   &  0.462789  \\
\end{tabular}
\end{ruledtabular}
\label{ex-paras}
\end{table}
%
\begin{figure}
\includegraphics*[width=9.0cm, height=10.0cm]{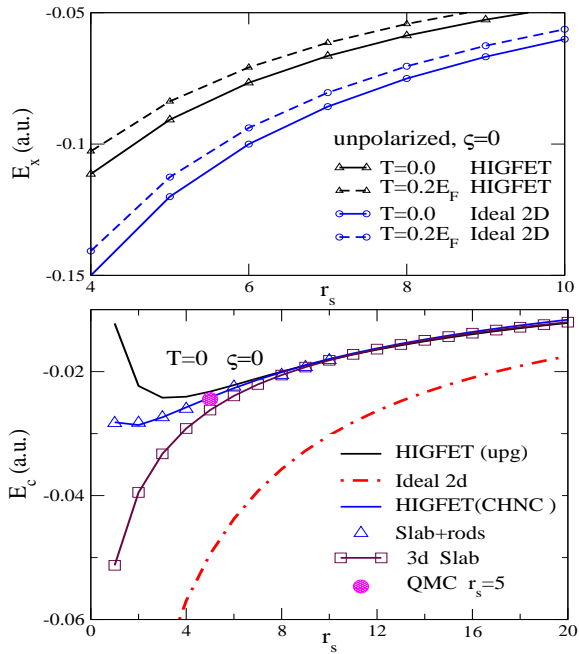}
\caption
{
(a)The exchange energy $E_x$ (Hartree a.u.) 
of a 2DES in a HIGFET compared
to that of an ideal 2DES, at $T/E_F=0$ and 0.2, with $\zeta=0$.
(b)The correlation energy $E_c$ at $T=0$ and 
$\zeta=0$, for the HIGFET layer. HIGFET(upg) is the
 ``unperturbed-$g$'' approximation. HIGFET(CHNC) is the full
calculation. This is  compared with the
correlation energy of a 3d slab model, and the ``slab+rod''model.
 The QMC datum for a
HIGFET ar $r_s=5$ is from De Palo et al\cite{depalo}.
}
\label{execfig}
\end{figure}
%
The exchange energy for a HIGFET with $n_d=0$ is shown in Fig.~\ref{execfig}
as a function of $r_s$ for $\zeta=0$ and  temperature $T/E_F$=0 and 0.2.
\section{The exchange-correlation energy for quasi-2D layers.}
\label{excpara}
 The correlation function $h^0(r)$ yields exact
 exchange energies for arbitrary layer
thicknesses.
 In contrast, the correlation energies
 require a coupling-constant
integration over  the pair functions $g(r,\zeta,w,\lambda)$
 calculated using the
 quasi-2D potential $\lambda W(r)$ for each $\lambda$.
 These  $g(r,\zeta,w)$
can be calculated using the CHNC. On the other hand,
the  {\it unperturbed}-$g$ 
approximation,
found to be useful in Quantum Hall effect studies\cite{ahm}, has been
exploited by
 De Palo
  et al.\cite{morsen}.
 They have used the pair functions $g(r,\zeta,w=0)$
 of the ideally thin layer obtained form QMC
 to calculate a correction energy
  $\Delta$ given by,
  \begin{equation}
  \label{unpertg}
  \Delta E =(n/2)\int 2\pi rdr [W(r)-V(r)]h(r,\zeta,w=0)
  \end{equation}
  Then the total exchange-correlation energy $E_{xc}(r_s,\zeta,w)$ is
  obtained by adding to $\Delta$ 
  the known $E_{xc}$ of the ideally-thin  system. 
 The above equation
  can also be applied at finite temperatures using the finite-$T$ pair
  functions $g(r_s,\zeta,T)$ obtainable from the CHNC procedure.   
 
 De Palo et al\cite{morsen} have performed Diffusion-Monte-Carlo 
 simulations at $r_s$ =5 for HIGFETS with $b=0.8707$, i.e,
 a CDM width  $w$=6.1256 a.u., and find that
the error in this approach compared to the full simulation is about
2\%. 
This full QMC result at $r_s=5$ is shown in the lower panel
of Fig.~\ref{execfig}.
Since the HIGFET system approximates to a thin-layer as $r_s$
increases, this approach is probably satisfactory for $r_s\ge $ 5. The
method becomes unreliable for small $r_s$, and definitely fails below $r_s=2$.
Also, the ``unperturbed-$g$'' approximation fails to include the
renormalization of the kinetic energy  picked up via the
coupling constant integration over the fully consistent $g(r,w)$.
 We report results(Fig.~\ref{execfig}) from
 the full coupling-constant integration
of the $g(r,w)$, (Fig.~\ref{execfig},
 lower panel, CHNC)
 as well as from the ``unperturbed-$g$'' 
 approximation\cite{morsen} used by De Palo et al.

	In parametrizing the quasi-2D
 correlation energy $E_c(r_s,w)$, we present an intuitive model
 of $E_c(r_s,w)$. For small $r_s$,
the ratio $w/r_s$ is large and the electrons are like 1-D wires with
the axis normal to the 2D plane and interacting with a $\log(w/r)$ 
interaction (c.f., Eq.~\ref{cdmeqn}). However, at large $r_s$ we have
3-D like electron disks with $w$ and $r_s$ of comparable magnitude in 
the density regimes of interest in HIGFETS. Thus we model the
quasi-2D $E_c$ as an interpolation between a 1D like 
form  and a 3D like form. First we consider a purely 3D model.
 Given the 2D-density $r_s$ and an effective CDM width $w$,
 we define an {\it effective} 3D 
 density parameter $r_s^{3d}$,
 purely for 
 calculating its correlation energy. It will be 
 seen that this 3D model is excellent for $r_s>7$. When
 $r_s$ becomes small (i.e, less than $\sim 3$),
 the effective width of the 2D layer,
 viz., $w/r_s$ becomes large and a 1D log-interaction
 model\cite{samaj} is needed.  
  To capture the rod-like regime,
 we define the ``rod like'' correction $\Delta E_c$ for $r_s<7$ by:
\begin{equation}
 \label{2drods}
 \Delta E_c(\zeta=0)=a_0+a_L\log(1/r_s)+a_1r_s
 \end{equation}
 where, for HIGFETS, $a_L=0.0221788$, $a_1=0.00365169$, and $a_0=0.0192979$
  for $\zeta=0$. This $\Delta E_c$ is added to the 3-D slab form 
  given below.
  The fit parameters for the $\zeta=1$
  ``rod-like'' correction are $a_0$=0.013337, $a_L$=0.0084787,
  and $a_1$=0.0006821, to be applied for $r_s<15$.
  
 For the 3D slab-like regime (i.e, $r_s>7$ for $\zeta$=0, 
 $r_s>15$ for $\zeta=1$) we
 define a $\zeta$-dependent 3-D density parameter and a
 correlation energy via: 
 \begin{eqnarray}
 \label{3dslab}
 E_c(r_s,\zeta,w)&=&E^{3d}_c(r_s^{3d},\zeta) \\
 R_s&=&r_s/F(r_s)\\
 r_s^{3d}&=&[wR_s^2]^{1/3}Z(\zeta)\\
 Z(\zeta)&=&\frac{2\surd{2}}{(4-\zeta)^{1/2}}(c_1^{1.5}+c_2^{1.5})
 \end{eqnarray}
 The 3D correlation energy $E^{3d}_c(r_s^{3d},\zeta)$ is that given by,
 e.g., Ceperley and Alder\cite{cepalder},
 or Gori-Giorgi and Perdew\cite{g-gp}.
  We see from the lower panel of
 Fig.~\ref{execfig} that the correlation energy for small $r_s$, 
 calculated using the 3D slab begins to
 go below the ``unperturbed-g'' approximation
 of Eq.~\ref{unpertg}, consistent
 with the trend of the ideal 2D gas, and the trend of the only
  QMC data point available for a HIGFET, at $r_s=5$. The exchange-correlation
energy  obtained from the full CHNC calculation is in excellent agreement
with the QMC datum. The curve labeled  "Slab+rods" 
in Fig.~\ref{execfig}  is the combined formula,
Eq.~\ref{2drods}, and Eq.~\ref{3dslab}, for the correlation energy/electron,
with, for example, the region $r_s\sim 7$ for
 $\zeta=0$ obtained by linear interpolation
between $r_s$=6 and $r_s=8$.
This is clearly seen to
reproduce the full CHNC results very well. 
 \subsection{Correlation energy at finite temperatures.}
 The correlation contribution to the Helmholtz free energy of the
 ideal 2-D layer, and layers of thickness $w$ can be easily calculated
 using the approach of Eq.~\ref{unpertg}, where the CHNC-generated
 finite-$T$ pair functions are used. A typical set of results
 at very low temperatures
 is given in Table~\ref{fc_table}. Here we have also given the
 packing fraction $\eta$ of the hard-sphere bridge function used
 to mimic the three-body and multi-body correlation contributions.
 As discussed in earlier work\cite{eplmass},
  the $F_x$ and the $F_c$ at very
  low $T$ contain logarithmic terms which cancel with
  each other, so that the sum $F_{xc}=F_x+F_c$ is free of such
  terms. From our numerical work we find that the $T$ dependence
  of the $F_x$ and $F_c$ of layers of finite thickness is very similar
  to that of ideally thin layers. Hence we assume that the
  logarithmic corrections are also similar. 
  At $r_s=5$ the cancellation is good to about 75\%, and
  this improves as $r_s$ increases.
 Although the two-component fluid (up spins
  and down spins) involves three distribution functions, we have,
  as before\cite{prl2}, used only one hard-disk bridge function, $B_{12}$, as
  clustering effects in $g_{ii}$ are suppressed by the Pauli-exclusion. 
  However, at high densities (low $r_s$), the use of three bridge functions
  seems to be needed for satisfying various subtle features that
  are needed to ensure the exact cancellation of logarithmic energy terms
  etc. Instead of introducing additional features
 into the CHNC method, we have however
  retained the single bridge-function model that
 was used by us so far\cite{prl2}.
\begin{table}
\caption{Low-temperature data for
the exchange-correlation contribution to the Helmholtz free energy,
$F_{xc}$ per electron in
atomic units at $r_s=5$, for the ideal 2D system and 
for a CDM of width 6.1256 au., i.e., Fang-Howard $b$=0.8707.
The packing fractioin $\eta$ which defines the hard-disk bridge
function is also given}
\begin{ruledtabular}
\begin{tabular}{cccccc}
$t=T/E_F$& $\eta$&$ F_{xc}$(ideal)& $F_{xc}(Eq.~\ref{unpertg})$&$F_{xc}$(CHNC)
\\
\hline
0.05 & 0.3718  &  -0.16902 & -0.11197 &-0.11467\\
0.08 & 0.3725  &  -0.16883 & -0.11177 &-0.11448\\
0.10 & 0.3730  &  -0.16867 & -0.11161 &-0.11433\\
0.12 & 0.3733  &  -0.16850 & -0.11144 &-0.11417\\
\end{tabular}
\end{ruledtabular}
\label{fc_table}
\end{table}
\begin{figure}
\includegraphics*[width=9.0cm, height=10.0cm]{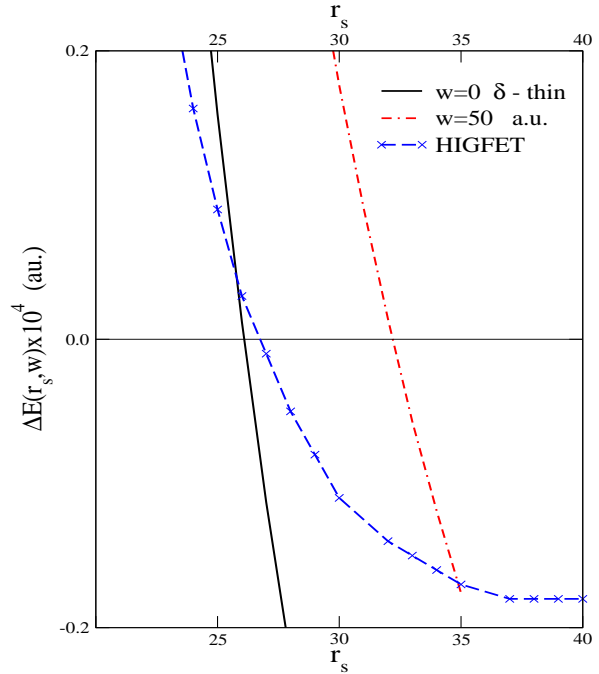}
\caption
{The Energy difference between the fully polarized and unpolarized
states of 2D layers  with different CDM widths $w$. 
The HIGFET $w$ varies as $r_s^{2/3}$
and is $\sim$ 18 near the spin-phase transition.
}
\label{sptfig}
\end{figure}
%
\subsection{The transition to a spin-polarized phase.}
QMC simulations as well as CHNC calculations show that
there is a spin-polarization transition (SPT) in the ideally-thin 2D
electron fluid near $r_s\sim 26$. On the other hand, the correlation
contributions dominate over the exchange energy in the 2-valley 2D system
in Si MOSFETS, and the SPT is suppressed\cite{2valley}. The rapid
increase in $m^*$, with $g^*$ remaining unchanged while $r_s$ is increased,
observed by Shashkin et al.\cite{shash} was found to be consistent
with this picture\cite{eplmass}. In finite-thickness
2D layers, as the CDM width $w$ increases, the location of the
SPT is pushed to higher values, as seen in 
Fig.~\ref{sptfig}.
In the case of HIGFETS used by, e.g., Tan et al.\cite{tan}, 
the width $w$ increases with $r_s$, but only as
$r_s^{2/3}$. Thus at $r_s$=26, the  $w$ is only 18.4, and the SPT 
remains intact and occurs at a somewhat higher $r_s$, as shown by de Palo
et al\cite{morsen}, and also in Fig.~\ref{sptfig}. 
A natural consequence of delaying the
SPT is to decrease the spin-susceptibility enhancement.
The effective thickness of the quasi-2D layer can be
increased by suitably designing the shape of the potential well, or
including an additional subband, and in this case the 
SPT can be circumvented. However, a discussion of higher subband effects
is outside the scope of this study.
\begin{figure}
\includegraphics*[width=9.0cm, height=10.0cm]{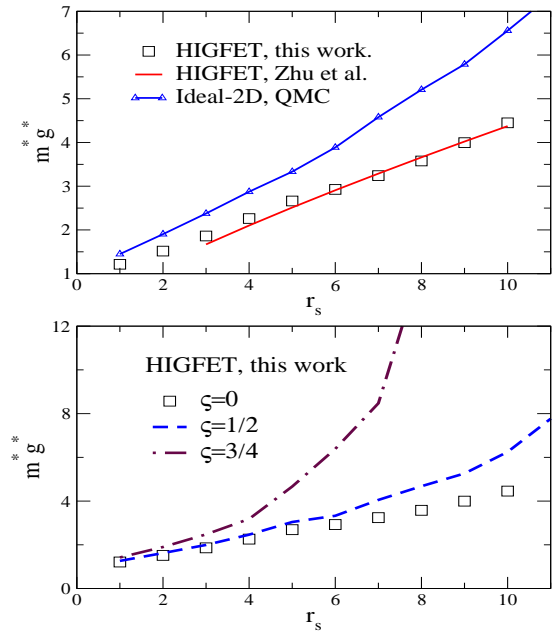}
\caption
{(a)The spin-susceptibility enhancement, i.e., $\chi_s/\chi_P$=$m^*g^*$ 
as a function of
$r_s$ for the HIGFET. The experimental data of Zhu
et al.\cite{zhu} are also shown.
(b)Variation of $m^*g^*$ as a function of the spin-polarization for
a HIGFET.
}
\label{susfig}
\end{figure}
\section{The spin-susceptibility, effective mass and the $g$-factor}
The results for the exchange-correlation free energy $F_{xc}(rs,\zeta,T)$
for the ideal 2D system and the thick-layer system contain all the
information needed to calculate the spin-susceptibility enhancement,
the effective mass $m^*$ and the effective Land\'{e} factor $g^*$,
for the ideal system and the thick  layer.
In fact, the following quantities are calculated from the
respective second derivatives of the energy.
\begin{eqnarray}
m^*&=&C_v/C_v^0=1+\frac{\left[\partial^2 F_{xc}(t)/\partial t^2\right]}
{\left[\partial^2 F_0(t)/\partial t^2\right]} \\
\chi_P/\chi_s&=& (m^*g^*)^{-1} = 1+
\frac{ \left[\partial^2 F_{xc}(\zeta)/\partial \zeta^2\right]}
 {\left[\partial^2 F_0(\zeta)/\partial \zeta^2\right]}
 \end{eqnarray}
 We use the available QMC results for the
 ideal 2D exchange-correlation energy at $T=0$, and where convenient,
 the QMC pair-distribution functions at $T=0$ as parametrized by Giri-Giorgi
 et al\cite{ggpair}. The CHNC is used to obtain the pair-functions for 
 situations (e.g, at finite-$T$ and finite thickness) 
 where the QMC data are simply not available or difficult to use.
 In most cases, replacing the QMC-PDFs with the CHNC ones, or
 using the ``unperturbed-$g$'' approximation leads to relatively
  small changes. The exception is in the calculation of $m^*$,
 where the ``unperturbed-$g$'' approximation, Eq.~\ref{unpertg},
 is inadequate. 
 \subsection{The effective mass $m^*$.}
\begin{figure}
\includegraphics*[width=9.0cm, height=10.0cm]{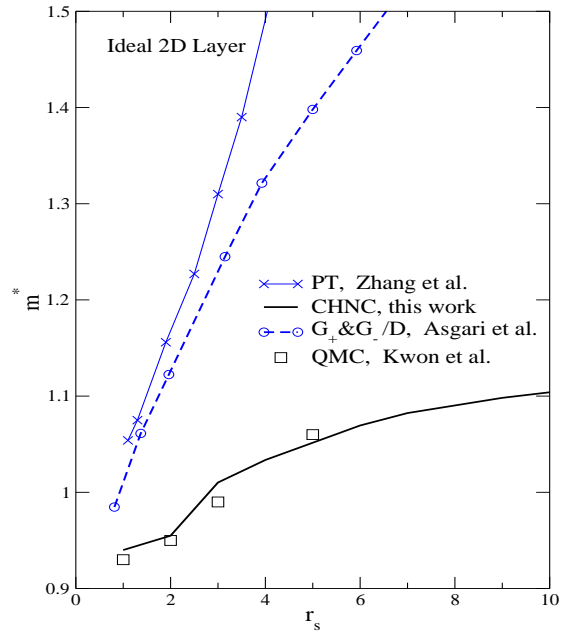}
\caption
{The effective mass $m^*$ of an ideal 2D layer ($w=0$)
 obtained from CHNC are
 compared with the the Quantum Monte Carlo data of Ref.~\cite{kwon} and the
perturbation theory calculations of Zhang et al.\cite{zhang},
and Asgari et al.\cite{asgari}, i.e., their
$G_+\&G_-/D$ calculation. 
}
\label{m2d-id}
\end{figure}

 In  Fig.~\ref{m2d-id} we present our results for the ideal-2DES. No
experimental results are available for this case, but limited QMC
simulations\cite{kwon}
 as well as results from diagrammatic perturbation theory (PT)
\cite{asgari,zhang} are available. 
The CHNC based $m^*$
 values have an error of at most $\pm$ 2\%. The ideal 2D-layer $m^*$ 
 shows a rapid rise around
 $r_s$=2 to 5, in good agreement with the four values from QMC,
 and then slows down in strong contrast to the $m^*$ proposed
 by Asgari et al., and Zhang et al. We have displayed the model denoted 
 $G_+\&G_-/D$ by Asgari et al., as being their optimal choice from among
 the many models given in Ref.~\cite{asgari}, where they also
 contest the analysis of Zhang et al. The PT approaches are
 partly semi-phenomenological in that QMC data are input into
 local-field factors and other ingredients of the calculation; 
 the choice of the vertex functions, treatment of on-shell or
 off-shell effects, whether to use self-consistent propagators or not, 
 etc., are components of the prescription used by different workers.
The strong disagreement of the PT-based $m^*$ with the QMC-based $m^*$
is notable.

The CHNC method has some similar uncertainties, especially in the
use of a Percus-Yevik hard-sphere bridge function $B(r)$ to capture the
back-flow like three-body contributions to the PDFs and the total
energy. As seen in Fig.~\ref{bridge}, the $T$ dependence
of $B(r)$ seems quite small, and our initial calculations of
$m^*$, reported in Ref.~\cite{ssc} were based on the zero-$T$
form of $B(r)$. This leads to an $m^*$ which drops slightly below unity
and remains there. In this calculation we have used the
proper $T$-dependent bridge function
and the calculated $m^*$ is in good agreement with the QMC-based
$m^*$. This might be somewhat  coincidental, as the
QMC results are also based on sensitive approximations. However, it
means that we do have a $B(r)$ which is consistent
with current QMC results, and hence may be used with 
greater confidence in studying thick-2D systems. Another
point in favour of our model of $B(r)$ is seen in the local-field
factor (LFF) of the ideal 2DES response function.
 A study of the LFF of the 
 2D response\cite{locf}
 shows that the formation of singlet-pair correlations is essentially
 complete by $r_s\sim 5$, and after that the structure of the fluid
 remains more or less unchanged, until the SPT is reached. The  
hard-sphere model of $B(r)$ provided a satisfactory description
of the short-ranged features of the 2DES-LFF.
The
 rapid  rise in $m^*$
 up to $r_s\sim 5$ and the subsequent slow-down is probably related to 
 the formation and persistence of the singlet structure in the 2D fluid
revealed by the form of the LFF.
\begin{figure}
\includegraphics*[width=9.0cm, height=10.0cm]{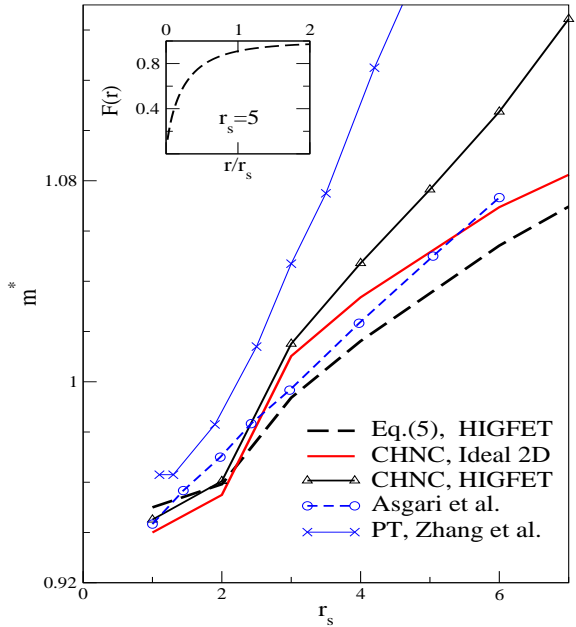}
\caption
{The effective mass $m^*_H$ in a HIGFET. The PT results of Zhang et al.,
and Asgari et al., for the HIGFET are  close to the ideal-2D results
from QMC. The ``unperturbed-$g$'' approach to the energy using CHNC-based
PDFs gives a slight reduction from the ideal 2D. For the full CHNC calculation,
see Fig.~\ref{scfmstr}}
\label{m2d-qua}
\end{figure}

 In  Fig.~\ref{m2d-qua} we present a comparison of various theoretical
calculations  of the 
 effective mass $m^*_H$ of the electrons in the HIGFET.  The PT
calculations of Zhang et al., and Asgari et al., show a strong decrease
of $m^*$ from the PT values in the ideal 2DES. Our calculations,
using the ``unperturbed-$g$'' approximation, Eq.~\ref{unpertg}, lead
to a $m^*_H$ which is only slightly reduced by the thickness
 effect. This $m^*_H$ curve is in close to that of
 Asgari et al. This is clearly a numerical accident.
 According to Asgari et. al., the difference between the ideal $m^*$ and
 $m^*_H$ increase as
 $r_s$ increases. In our calculation using the ``unperturbed-$g$''
 approximation, the difference, already quite small,
  seems to diminish as $r_s$ increases. In fact, as $r_s$ increases, the
  ratio $w/r_s$ of the HIGFET layer decreases and the thickness effect
  may be expected to decrease, unless the difference
   between $m^*$ and $m^*_H$ is driven by some other effect.
  This ``other effect'' is revealed by giving up the ``unperturbed-$g$''
  approximation, and using the full thick-layer 2DES pair-distribution
 function at finite $T$, calculated using the CHNC, to evaluate 
 the total free energy $F(r_s,T)$ of the quasi-2D system, and hence the
$m^*_H$. 
In Fig.~\ref{scfmstr}
 we display the PDF of the quasi-2DES of a HIGFET at $r_s=5$, and
 compare it with the PDF of the ideal 2DES. The difference between
 the ideal and quasi systems is embodied in the form factor $F(r)$.
 The reduced Coulomb repulsion at small-$r$ leads to {\it a large pile-up
 of electrons around the electron at the origin. } This means, the
electron has to drag this charge pileup and this contributes an
enhanced $m^*$ to the
 thermodynamic and transport properties of the quasi 2DES.

The experimental results of Tan et. al.,
 for $m^*$ show an increase of $\sim$ 150\%
  between 
$r_s$=3 and $r_s=6$. Our results from the full CHNC calculation, as 
well as the perturbation results of Asgari et al., are shown
in the lower panel of  Fig.~\ref{scfmstr}. Given the failure of
the PT calculations to reproduce the QMC-based $m^*$ for the ideal 2D,
it is difficult to evaluate the reliability of the PT-based $m^*_H$.
The PT-overestimate of $m^*$ of the ideal 2D 
is typical of the RPA-like character
of these theories which are likely to predict spin transitions
at relatively high densities.
Also, we believe that if the same PT prescriptions were used to
evaluate the one-top value $g(0)$ of the PDFS of the 
2DES and the quasi-2DES,
another measure of the short-comings of the PT methods would be revealed.

\subsection{Enhanced spin susceptibility and the Lande-$g$ factor.}
\begin{figure}
\includegraphics*[width=9.0cm, height=11.0cm]{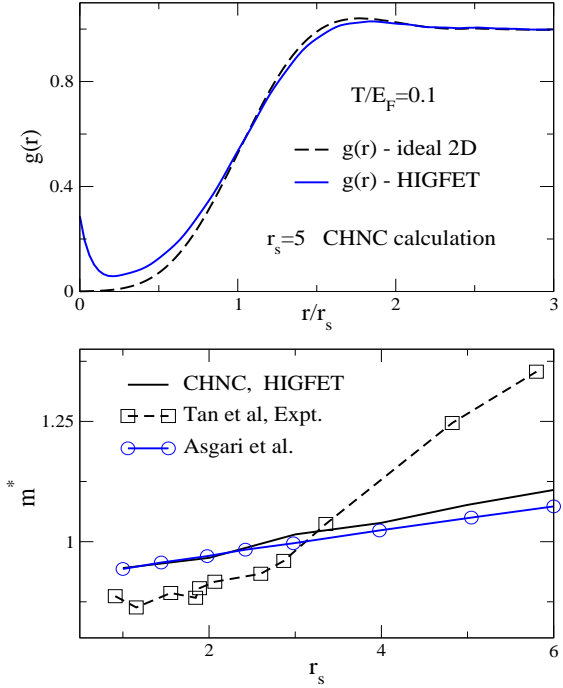}
\caption
{(a)The spin-averaged $g(r)$ at $r_s=5$
and $T/E_F$=0.1,
for an ideal 2DES, and for a HIGFET  calculated using CHNC.
(b)The effective mass $m^*_H$ in a HIGFET from fully self-consistent
CHNC calculations finite-$T$.
 The experimental data are from
 Tan et al\cite{tan}.
}
\label{scfmstr}
\end{figure}

\begin{figure}
\includegraphics*[width=8.0cm, height=9.0cm]{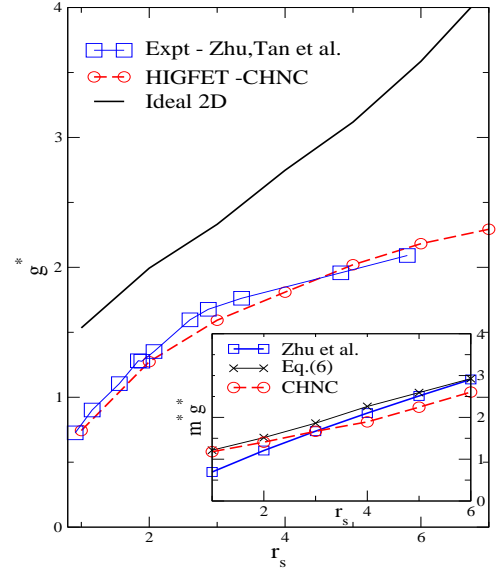}
\caption
 { The ideal 2DES $g$-factor is obtained
from the QMC $m^*g^*$ of ref.~\cite{attac}, divided by the CHNC ideal 2D $m^*$.
The experimental HIGFET $g_H^*$ (boxes) is from the $m^*g^*$ of Zhu et al.,
divided by the $m^*$ of Tan et al.
The inset shows the spin-susceptibility enhancement $m^*g^*$ from the
$E_{xc}(r_s,\zeta)$ calculated from Eq.~\ref{unpertg}, where
the ideal 2D $g(r,w=0)$ is used, and
from the full CHNC calculation using the $g(r,w)$ consistent with
the quasi-2D potential.
}
\label{gstar}
\end{figure}

The product $m^*g^*$ is given by the 
ratio of the static spin susceptibility $\chi_s$  to
 the ideal (Pauli) spin susceptibility $\chi_P$.
The long wavelength limit of the static
response functions are connected with the
compressibility or the spin-stiffness via the second derivative of
the total energy with respect to the density
 or the spin polarization\cite{2valley}.
De Palo et al.\cite{morsen} have calculated $m^*g^*$ from the 
QMC pair distribution functions and shown that they obtain
quantitative agreement with the data for
 very narrow 2-D systems\cite{vakili}
as well as for the thicker systems found in HIGFETS\cite{zhu}.
The CHNC PDFs are  close approximations to
QMC results, and when used in Eq.~\ref{unpertg}, yield correction
energies which are in good agreement with 
the energies obtained
 by De Palo et al\cite{depalo}. 
In Ref.~\cite{2valley} we showed that the rapid enhancement of $m^*g^*$
in Si/SiO$_2$ 2DES 
is a consequence of the increase in $m^*$ with $r_s$,
 and that the $g^*$ does
{\t not} increase with $r_s$
 because there is {\it no spin-phase transition} in the
2-valley case. In the HIGFET system there is a slightly delayed SPT,
as seen in Fig.~\ref{susfig}.
Hence $m^*g^*$ increases with $r_s$, while $m^*$ also increases
quite rapidly, due to the enhanced ``on-top'' correlations 
shown (Fig.~\ref{scfmstr}) in
the PDF of the quasi-2DES. The resulting $g^*$ of the HIGFET is shown
in Fig.~\ref{gstar}.
\section{Conclusion.}
We have presented a detailed study of 
the effect of many-body interactions
in quasi-2D electron layers using a single theoretical framework which
involves the calculation of the pair-distribution functions of the system via a
classical representation of the quantum fluid. A procedure for replacing the
inhomogeneous transverse distributions via a constant-density model, i.e., an
equivalent {\it homogeneous} distribution, has also been demonstrated.
Easy to use parametrized fit formulae for the exchange energy at zero and
finite-$T$ have been presented. A simple numerical scheme for calculating the
correlation energy of a thick 2D layer,, {\it via} a 3-D slab model
combined with a 1-D rod model, has also been demonstrated. 
We find that the thickness effect 
on the spin-phase transition etc., provides a clear picture of the 
changes in the spin-susceptibility enhancement leading to a strong
increase in the $g$-factor, while $m^*$ is increased due to the
enhancement of the ``on-top'' correlations arising from the
reduction of the Coulomb potential in thick layers.
 However, unlike in the case of the effective mass data
for Si/SiO$_2$ systems\cite{shash,eplmass},
these results do not provide good
 quantitative agreement with the effective-mass data
for HIGFETS recently reported by Tan et al. This may be
due to our use of the ideal 2DES bridge function even for
the HIGFETS.

\end{document}